\documentclass[11pt]{article}
\usepackage{amsmath,amssymb,epsfig,sint}
\font\sixrm=cmr6

\newcommand{\R}{\rm I\kern-.2emR}
\newcommand{\C}{\rm \kern.25em\vrule height1.4ex
depth-.12ex width.06em\kern-.31em C}
\newcommand{\N}{{\rm I\kern-.16em N}}
\newcommand{\Z}{{\rm Z\kern-.35em Z}}

\newcommand{\cl}{{\ell}}

\newcommand{\kappac}{\kappa_{\rm c}}

\newcommand{\gr}{g_{{\hbox{\sixrm R}}}}
\newcommand{\phir}{\phi_{{\hbox{\sixrm R}}}}
\newcommand{\mr}{m_{{\hbox{\sixrm R}}}}
\newcommand{\vr}{v_{{\hbox{\sixrm R}}}}
\newcommand{\Zr}{Z_{{\hbox{\sixrm R}}}}
\newcommand{\Zrhat}{\widehat{\Zr}}

\newcommand{\rme}{{\rm e}}

\newcommand{\rmO}{{\rm O}}

\newcommand{\be}{\begin{equation}}   
\newcommand{\ex}{\end{equation}}
\newcommand{\ba}{\begin{eqnarray}}
\newcommand{\ea}{\end{eqnarray}}

\newcommand{\bfx}{{\bf x}}

\newcounter{subequation}[equation]
\makeatletter

\expandafter\let\expandafter
\reset@font\csname reset@font\endcsname

\def\subeqnarray{\arraycolsep1pt
    \def\@eqnnum\stepcounter##1{\stepcounter{subequation}%
        {\reset@font\rm(\theequation\alph{subequation})}}
\jot5mm     \eqnarray}

\makeatother
\newcommand{\msbar}{{\rm \overline{MS\kern-0.14em}\kern0.14em}}

\begin{document}
\begin{titlepage}

\begin{flushright}
   MPP-2004-152\\
   December 2004
\end{flushright}

\vskip 0.20 true cm

%\vskip 0.20 true cm
\begin{center}
{\Large\bf 
The $4d$ one component lattice $\phi^4$ model in the broken phase revisited}
\end{center}
\vskip 1 true cm
\centerline{\large Janos Balog}
\vskip1ex
\centerline{Research Institute for Particle and Nuclear Physics}
\centerline{1525 Budapest 114, Pf. 49, Hungary}
\vskip 1 true cm
\centerline{\large Anthony Duncan, Ray Willey}
\vskip1ex
\centerline{Department of Physics, 100 Allen Hall}
\centerline{University of Pittsburgh, Pittsburgh, PA 15260, USA}

\vskip 1 true cm
\centerline{\large Ferenc Niedermayer${}^*$}
\vskip1ex
\centerline{Institute for Theoretical Physics, University of Bern}
\centerline{CH-3012 Bern, Switzerland}
\vskip 1 true cm
\centerline{\large Peter Weisz}
\vskip1ex
\centerline{Max-Planck-Institut f\"ur Physik}
\centerline{F\"ohringer Ring 6, D-80805 M\"unchen, Germany}
\vskip 1 true cm
\centerline{\bf Abstract}
\vskip 1.0ex
Measurements of various physical quantities in the symmetry broken
phase of the one component lattice $\phi^4_4$ with standard action,
are shown to
be consistent with the critical behavior obtained by renormalization
group analyses. 
This is in contrast to recent conclusions by another group, who further 
claim that the unconventional scaling behavior they observe, 
when extended to the complete
Higgs sector of the Standard Model, would alter the conventional 
triviality bound on the mass of the Higgs.

\vfill
\noindent{---------------}\\
\noindent\footnotesize{${}^*$On leave from E\"otv\"os University, 
HAS Research Group, Budapest, Hungary} 
\eject

\end{titlepage}

\section{Introduction}

The continuum limit of the lattice regularized $\phi^4$ theory in 
4 dimensions with $n\ge1$ components is thought to be trivial.
The evidence comes from renormalization group studies and numerical
simulations; remarkably however there is still no rigorous proof.

After a period of intense numerical investigations about 15 years ago,
interest in this model within the lattice community dwindled.
Occasionally however some authors have questioned 
the usual procedures of renormalization
in the broken phase. In their latest paper \cite{CCC} 
Cea, Consoli and Cosmai claim to present further numerical evidence 
for their scenario. 
From the presentation of the results in their paper a neutral reader
would, accepting their analysis, indeed conclude that the
conventional wisdom (CW) concerning the critical behavior in the 
lattice $\phi^4$ was incorrect. 
The purpose of this paper is to point out the
deficiencies of their analyses, and to show that the numerical data is 
completely consistent with the conventional picture.

\section{The lattice theory and renormalization group predictions}

The standard lattice $\phi^4$ model is characterized by two bare 
parameters $\kappa,\lambda$; the action is
\footnote{We use the notations in refs.~\cite{LWsymm,LWbroken}.} 
\be
S=\sum_x\left[-2\kappa\sum_{\mu=1}^4\phi(x)\phi(x+\hat{\mu})+\phi(x)^2
+\lambda(\phi(x)^2-1)^2\right]\,.
\end{equation}
There are two phases separated by a line of 
critical points $\kappa=\kappac(\lambda)$. For $\kappa>\kappac(\lambda)$
the symmetry is spontaneously broken and the bare field $\phi(x)$ 
has a non-vanishing expectation value $v$. As usual this is defined by
first taking the thermodynamic limit in the presence of an external 
uniform magnetic field $h$ and then letting this field tend to zero
\be
v=\lim_{h\to0}\lim_{V\to\infty}\langle\phi(0)\rangle_h\,.
\label{vevh}
\end{equation}

This study will restrict attention to the vacuum expectation 
value and the connected two-point function
\be
G(x)=\langle\phi(x)\phi(0)\rangle_{\rm c}=
\langle\phi(x)\phi(0)\rangle-v^2\,.
\end{equation}
A renormalized mass $\mr$ and a field wave function renormalization
constant $\Zr$ are defined through the behavior of Fourier transform of 
$G(x)$ for small momenta:
\be \label{eq:tGk}
\tilde{G}(k)^{-1}=\Zr^{-1}\left\{\mr^2+k^2+\rmO(k^4)\right\}\,.
\end{equation}
The susceptibility $\chi$ and the second moment $\mu$ are defined through
\ba \label{eq:susc}
\chi&=&\sum_x G(x)=\Zr/\mr^2\,,
\label{chi}
\\
\mu&=&\sum_x\,x^2G(x)=8\Zr/\mr^4\,.
\ea
We also define the normalization constant associated with the
canonical bare field through
\footnote{In various papers a different notation is employed
and the quantity $16\kappa\chi^2/\mu$ is denoted by $\Zr$.}
\be
\Zrhat=2\kappa\Zr=2\kappa\mr^2\chi\,.
\end{equation}
In the framework of perturbation theory correlation functions of
the multiplicatively renormalized field 
\be
\phir(x)=\Zr^{-1/2}\phi(x)\,,
\end{equation} 
have at all orders finite continuum limits after mass and coupling 
renormalization are taken into account.
Correspondingly a renormalized vacuum expectation value is 
defined through
\be
\vr=v\Zr^{-1/2}\,.
\end{equation}
Here we adopt the generally accepted  
assumption that the structurally same multiplicative 
renormalization is required to define the continuum 
limit of the theory non-perturbatively.
We see no evidence for the claim in ref.~\cite{CCC} 
that the vacuum expectation value and the fluctuating part 
of the bare field should be renormalized differently. 
Finally a particular renormalized coupling is defined by 
\be
\gr\equiv 3\mr^2/\vr^2=3\mr^4\chi/v^2\,,
\label{gr}
\end{equation}
as is one popular choice in the perturbative framework.

The commonly accepted picture of the critical behavior of the theory 
is mainly due to the work of the Saclay group \cite{BGZ}. In their
framework the renormalization group equations predict that
the mass and vacuum expectation value go to zero according to
\ba
\mr&\propto&\tau^{1/2}|\ln(\tau)|^{-1/6}\,,
\\
\vr&\propto&\tau^{1/2}|\ln(\tau)|^{1/3}\,,
\ea
for $\tau=\kappa/\kappac-1\to0$,
and correspondingly the renormalized coupling is predicted 
to go to zero logarithmically which is the expression of triviality.
The critical behavior in the broken phase is conveniently expressed in 
terms of three integration constants $C'_i\,,i=1,2,3$ appearing in the 
critical behaviors:
\ba
\mr&=&C'_1(\beta_1\gr)^{17/27}\rme^{-1/\beta_1\gr}
\left\{1+\rmO(\gr)\right\}\,,\,\,\,\,\beta_1=\frac{3}{16\pi^2}\,,
\\
\Zr&=&C'_2\left\{1-\frac{7}{36}\frac{\gr}{16\pi^2}+\rmO(\gr^2)\right\}\,,
\label{zrrg}
\\
\kappa-\kappac&=&\frac12 C'_3\mr^2\gr^{-1/3}\left\{1+\rmO(\gr)\right\}\,.
\ea
In ref.~\cite{LWbroken} these constants were estimated by relating
them to the corresponding constants $C_i$ in the symmetric phase. These 
in turn were computed by integrating the renormalization group
equations with initial data on the line $\kappa=0.95\kappac(\lambda)$
obtained from high temperature expansions.

\section{Some proposed criticisms of CW}

In ref.~\cite{CCC} the authors present two quantitative arguments against
the conventional wisdom. Their work considers just the Ising limit
($\lambda=\infty$). Firstly they 
claim $\Zr$ grows logarithmically as one approaches the critical
point instead of going to a constant as predicted 
by the RG (see Eq.~\eqref{zrrg}).
Unfortunately they only measure $v,\chi$ which is not sufficient
to determine $\Zr$. To overcome this deficiency they compute
$\widetilde{\Zr}=m_{\rm input}^2\chi$, where $\chi$ is their MC
measurement at a given value of $\kappa$ but $m_{\rm input}$ is an 
estimate of $\mr$ at the same $\kappa$ taken from Table~3 of
ref.~\cite{LWbroken} (referred to as T3 below). 
This ``composed" quantity $\widetilde{\Zr}$ indeed
seems to grow as the critical point is approached. Accepting that
their measurements of $\chi$ are correct (and indeed we agree with
their values on the lattices we checked), the crucial question is whether 
the estimates $m_{\rm input}$ of $\mr$ are reliable.

At this stage two
details of the analysis in ref.~\cite{LWbroken} must be appreciated. 
Firstly the values of $\kappa$ cited in the last column of T3 
are written using $\kappac$ obtained from 
the same procedure of analysis for all values of $\lambda$ 
\footnote{In hindsight it would have been better to 
have tabulated estimated values of $\kappa-\kappac$}. 
For the particular case of the Ising model there 
are (presumingly) more accurate values; 
some comparisons are made in Table~\ref{kappac}. 
Hence, even if one 
accepts the estimated errors in T3, estimates of $\mr$ for a given 
$\kappa$ obtained using this table naively are subject to further large 
uncertainties.

\begin{table}[ht]
\centering
\begin{tabular}[t]{l|c}
\hline
$\,\,\,\,\,\,\,\kappac$&ref.\\[1.0ex]
\hline \hline
$0.07475(7)$&\cite{LWsymm}\\[1.0ex]
$0.074834(15)$&\cite{GSM}\\[1.0ex]
$0.074848(2)$&\cite{StaufAd}\\[1.0ex]
$0.074851(8)$&\cite{KL}\\[1.0ex]
$0.07487(5)$&\cite{VW}\\[1.0ex]
\hline
\end{tabular}
\caption{\footnotesize Estimates of $\kappac$ in the 4$d$ Ising model.}
\label{kappac}
\end{table}

Secondly it is probable that the systematic errors in T3
have been underestimated. 
This stems from a probable underestimate on the size of the
systematic errors on the cited values of the integration constants $C'_i$;
e.g. the estimated values of the constants obtained by using 
2-loop or 3-loop expressions in the RG equations differ considerably.
This question was briefly addressed in the 3rd paragraph sect.~5.2
of \cite{LWbroken} and later stressed by Peter Hasenfratz \cite{PH}
\footnote{This critique was discussed in more detail for the
physically relevant case of O(4) \cite{LWO4}, where fortunately the 
various loop estimates are closer.}. 

The simple outcome of the discussion above is that to compute $\Zr$ and 
$\mr$ independently one needs accurate {\it measurements of both} 
$\chi,\mu$. In the next sections we describe such measurements and 
find that the computed values of $\Zr$ are consistent with the RG 
expectation in the $\phi^4$ theory. For the Ising limit the 
measured values of $\mr$ are considerably lower than the corresponding 
estimates $m_{\rm input}$ in \cite{CCC}. 
 
The second presented ``evidence" against CW in \cite{CCC}
concerns the quantity $v^2\chi$. Defining
\be
\cl\equiv -\ln(\kappa-\kappac)\,,
\end{equation}
in the RG framework $v^2\chi$
behaves as $\propto\cl$ as one approaches $\kappac$,
whereas in their scenario they expect a more singular behavior 
$\propto\cl^2$. Figure~2 in \cite{CCC}
gives the impression that the data strongly supports their prediction.
However, this impression is completely false! The figure is obtained
by 2-parameter fits of $v^2\chi$ with functions $A\ln^p(\kappa-\kappac)$
with $p=1,2$ and $\kappac$ a fit parameter. 
But such fits are of course meaningless without including a scale
of the logarithm. Once this is done (as described in subsect.~4.4),
one gets a beautiful fit to the data also for the function
expected from the RG group.

\section{Ising MC simulation and results}

In this section we present our simulations of the 4$d$ Ising model
on a hypercubic lattice with periodic boundary conditions in all
directions. We consider only geometries with volumes $V=L^3T$.

\subsection{Simulation algorithm}

For updating the configurations we used the Swendsen-Wang 
cluster method \cite{SW}.
Beyond practically eliminating the problem of critical
slowing down, the method allows to use improved
estimators which reduce significantly the errors of the measured
quantities. 
Since the application of the cluster improved estimators 
have some specific features for the broken phase, we discuss
briefly here the applied procedure.

The first step in the SW cluster algorithm is to put
bonds between neighboring spins of equal signs with probability
$p=1-\exp(-4\kappa)$, while no bonds are put between
spins with opposite signs. In the second step one identifies
the clusters, the set of spins connected by bonds. 
Obviously, spins within a cluster have the same sign.
Denote the number of clusters by $\nu$, and the number of
sites in the $i$-th cluster by $n_i$, $i=1,\ldots,\nu$.
Assume for convenience that the numbering is chosen so that
$n_1 \ge n_2 \ge \ldots \ge n_\nu$.
In the updating step one flips all spins in the $i$-th cluster
with probability 1/2. All resulting $2^\nu$ configurations
appear in the equilibrium distribution with equal probability.
The cluster improved estimator replaces a quantity by its
average over these $2^\nu$ configurations.
In particular the product $\phi(x)\phi(y)$ is replaced by $+1$
if $x$ and $y$ belong to the same cluster, and by $0$ 
if the two sites belong to different clusters.

\subsection{Determination of $v$ and $\chi$}

The difference between the symmetric and broken phase shows
up in the distribution of the cluster size.
In the symmetric phase (at finite correlation length $\xi$)
the typical size of clusters is determined by $\xi$ and
remains finite and independent of $L$ for $L\gg \xi$.
This is obvious from considering the spin-spin correlator
$G(x)$ at large distances $|x|$
since this number is equal to the probability that the two
sites are within the same cluster.
As a consequence, this probability 
goes down exponentially with the distance.
It is easy to show that in the symmetric phase the susceptibility is
given by 
$\chi=\left\langle \frac{1}{V} \sum_i n_i^2 \right\rangle\,.$

In the broken phase (in infinite volume) 
$\langle\phi(x)\phi(y)\rangle\to v^2$ 
for $|x-y|\to\infty$. This involves that the size of the largest cluster
grows proportionally to the volume, 
$\langle n_1 \rangle \propto V$ while the distribution of 
$n_i$, $i=2,3,\ldots$ is not affected by the volume for $L,T\gg\xi$ 
(and is characterized by the correlation length).
In this case (for $V\to\infty$) the vacuum expectation value
is given by 
\be
v=\frac{1}{V}\langle n_1 \rangle\,,
\label{vevdef}
\end{equation}
and the susceptibility by
\be
\chi=\left\langle \frac{1}{V} \sum_{i>1}  n_i^2 \right\rangle +
\frac{1}{V}\left(\langle n_1^2\rangle -\langle n_1\rangle^2\right)\,.
\label{chiv}
\end{equation}
In sufficiently large volumes the definition Eq.~\eqref{vevdef}
of the vev coincides with the conventional definition Eq.~\eqref{vevh}.
For sufficiently large $V$ in the broken phase
one can choose an external field $h$ such that 
$h \langle n_1 \rangle = h v V \gg 1$ and at the same time
$h n_2 \ll 1$. Therefore in the presence of the magnetic field
the spins in the largest cluster are frozen to value $+1$ while
$h$ has practically no influence on the orientation
of the other clusters.
Eq.~\eqref{vevdef} provides a convenient definition of $v$ in 
a finite volume.
For large enough volumes this definition also coincides practically
with the Binder's definition \cite{Binder}
using the absolute value of the total magnetization:
\be
v_{\rm Binder}=\left\langle\frac{1}{V}|\sum_x\phi(x)|\right\rangle\,,
\end{equation}
which is the definition employed in ref.~\cite{CCC}.
For a thorough discussion of various finite volume definitions of the vev
(and more general finite volume effects in this model) we refer the
reader to ref.~\cite{JMMTW}.
Assigning random signs to the clusters $i=2,\ldots,\nu$ the sum of spins
in these clusters has a variance
$\sum_{i=2}^{\nu} n_i^2 \lesssim V\chi$ (cf. Eq.~\eqref{chiv}), while the 
value of the spin
in the largest cluster is $n_1 \sim v V$. The probability that the 
random component has the same magnitude as the constant component $n_1$
is suppressed at least by the factor $\exp(- V v/\chi^2)\,.$

Note that for the case of $V=L^3 T$, $L=\mathrm{fixed}$, $T\to\infty$ 
the situation is somewhat different.
In this quasi-one-dimensional case the large clusters
have finite length in the $t$-direction, although much larger than $L$. 
Their typical length is given by the inverse of the energy gap between the
lowest even and odd eigenstates of the transfer matrix \cite{JMMTW}. 
Here we shall not discuss this geometry, note only the obvious fact that 
any quantity defined through spin correlators can be expressed
in terms of expectation values of the cluster sizes $n_i$.

Apart from $v$ and $\chi$ we also measured the time-slice correlation
functions
\be
S(t)=\sum_{\bfx}G((t,\bfx))\,,
\label{timeslice}
\end{equation}
to compute the exponential mass 
and the Fourier transform $\tilde{G}(k)$ at momenta $k$ along the $t$-axis.

Typically we performed more than $400$K sweeps at each $\kappa$ point. 
The autocorrelation times for the vev and the two point function 
$S(t=T/4)$ were monitored and found reasonably small 
e.g. for the lattice with the
largest correlation length which we measured ($\kappa=0.0751$)
they were $\sim 8,\sim 3$ respectively.

\subsection{Raw data and analysis}

To illustrate the quality of the raw data, and for later reference,
in Table~\ref{rawdata0} the time-slice correlation 
function $S(t)$ is given for $V=48^4$ at $\kappa=0.0751$.

\begin{table}[ht]
\centering
\begin{tabular}[t]{r|l||r|l||r|l||r|l}
\hline
$t$ & \quad $S(t)$ & $t$ & \quad $S(t)$  & $t$ & \quad $S(t)$  & $t$ & \quad $S(t)$  \\
\hline
0 & 17.788(31) &  7 & 5.320(27) & 14 & 1.666(25) & 21  & 0.646(29)  \\
1 & 14.913(31) &  8 & 4.492(26) & 15 & 1.421(25) & 22  & 0.600(29)  \\
2 & 12.529(30) &  9 & 3.796(26) & 16 & 1.217(26) & 23  & 0.573(30)  \\
3 & 10.540(30) & 10 & 3.211(25) & 17 & 1.048(26) & 24  & 0.563(30)  \\
4 & $\phantom{1}$8.875(29) & 11 & 2.720(25) & 18 & 0.909(27) &     &  \\
5 & $\phantom{1}$7.478(28) & 12 & 2.306(25) & 19 & 0.798(27) &     &  \\
6 & $\phantom{1}$6.306(28) & 13 & 1.958(25) & 20 & 0.711(28) &     &  \\
\hline
\end{tabular}
\caption{{}\footnotesize The (subtracted) time-slice correlation function
$S(t)$ for the Ising case at $\kappa=0.0751$ on a $48^4$ lattice.
Note that the quoted errors for this quantity are much smaller than
that for the subtracted term $v^2L^3=2711.1(1.2)$.
%The subtraction for this quantity was $v^2 L^3=2711.099$.
%The quoted errors are for the subtracted quantity -- the error of 
%the constant part is much larger.
}
\label{rawdata0}
\end{table}

In Table~\ref{rawdata1} we collect data of the values of $v,\chi$
for various values of $\kappa,L$ from two papers \cite{JMMTW,CCC}
together with our own. The simulation at the largest value
of $\kappa=\kappa_0$ was performed to test our programs and analyses
by comparison at a point where the low temperature expansion
is considered quantitatively reliable \cite{VW}.
The agreement between the data and analytic computations is indeed
satisfactory. As for the values of $\kappa$ closer to $\kappac$, 
firstly one observes excellent agreement
of the raw data on those lattices measured by different groups.
Thus we are confident that the raw data (albeit sometimes obtained by
different methods) can be trusted. Secondly at values of $\kappa$ 
where different volumes were measured there are no signs of significant
finite volume effects.

\begin{table}[ht]
\centering
\begin{tabular}[t]{l|c|c|l|l|c}
\hline
$\,\,\,\,\,\kappa$&$L$&$\chi$&$\,\,\,\,\,\,\,\,v$
&$\,\,\,v^2\chi$&ref.\\[1.0ex]
\hline \hline
$\kappa_0$&$10$&$5.130(2)$&$0.571267(8)$&$1.674(1)$&\\[1.0ex]
\hline
$0.077$&$16$&$18.18(2)$&$0.38947(2)$&$2.758(4)$&\cite{JMMTW}\\[1.0ex]
\hline
$0.076$&$20$&$37.85(6)$&$0.30158(2)$&$3.442(6)$&\cite{JMMTW}\\[1.0ex]
$0.076$&$20$&$37.80(6)$&$0.30151(3)$&$3.437(6)$&\\[1.0ex]
$0.076*$&$20$&$37.84(6)$&$0.30158(2)$&$3.442(6)$&\\[1.0ex]
\hline
$0.0759$&$32$&$41.71(13)$&$0.29030(2)$&$3.515(12)$&\cite{CCC}\\[1.0ex]
$0.0759$&$48$&$41.95(93)$&$0.29028(5)$&$3.535(79)$&\cite{CCC}\\[1.0ex]
\hline
$0.075628$&$48$&$58.70(42)$&$0.25580(2)$&$3.841(28)$&\cite{CCC}\\[1.0ex]
\hline
$0.0754$&$32$&$87.45(76)$&$0.22054(8)$&$4.253(40)$&\cite{CCC}\\[1.0ex]
$0.0754$&$32$&$86.95(55)$&$0.22055(6)$&$4.229(29)$&\\[1.0ex]
$0.0754$&$48$&$87.82(56)$&$0.22048(2)$&$4.269(28)$&\cite{CCC}\\[1.0ex]
\hline
$0.075313$&$48$&$104.2(1.3)$&$0.20477(4)$&$4.367(56)$&\cite{CCC}\\[1.0ex]
\hline
$0.075231$&$60$&$130.8(1.4)$&$0.18812(3)$&$4.629(50)$&\cite{CCC}\\[1.0ex]
\hline
$0.0752$&$36$&$142.1(8)$&$0.18138(5)$&$4.674(29)$&\\[1.0ex]
$0.0752$&$48$&$142.6(9)$&$0.18132(4)$&$4.689(32)$&\\[1.0ex]
\hline
$0.0751$&$48$&$203.8(3.1)$&$0.15665(10)$&$5.002(82)$&\cite{CCC}\\[1.0ex]
$0.0751$&$48$&$206.4(1.2)$&$0.15657(4)$&$5.060(32)$&\\[1.0ex]
$0.0751$&$52$&$201.2(6.2)$&$0.15654(7)$&$4.93(15)$&\cite{CCC}\\[1.0ex]
$0.0751$&$60$&$202.4(8.6)$&$0.15648(2)$&$4.96(21)$&\cite{CCC}\\[1.0ex]
\hline
$0.074968$&$68$&$460.2(4.9)$&$0.11261(5)$&$5.836(67)$&\cite{CCC}\\[1.0ex]
\hline
$0.0749$&$68$&$1125(36)$&$0.07736(12)$&$6.73(24)$&\cite{CCC}\\[1.0ex]
$0.0749$&$72$&$1141(39)$&$0.07752(21)$&$6.86(27)$&\cite{CCC}\\[1.0ex]
\hline
\end{tabular}
\caption{\footnotesize Measured values of $\chi,v$ from various Ising 
simulations; data from this investigation have no entry in the last 
column. $\kappa_0=0.080795$.
In all cases $T=L$ except for the lattice where $\kappa=0.076^*$ denotes
$L=20, T=32$.}
\label{rawdata1}
\end{table}

The second moment mass can be determined directly from 
the Fourier transform of $S(t)$ (i.e. $\tilde{G}(k)$ with momenta along 
the $t$-axis) using Eq.~\eqref{eq:tGk} which involves 
a determination of the slope from the available discrete values of $k$.
To avoid possible discretization errors due to the finiteness of 
the lattice size $T$ we also used the following procedure
to determine $\mr$ and $\Zr$. We first computed the
exponential mass $m$ characterizing the exponential fall-off
of the time-slice correlation functions at large separations,
using one-mass and constrained two-mass fits (where the second mass
was required to satisfy $m_2\gtrsim 2m$). For the one-mass fit we 
included only distances $t$ with $mt\gtrsim1$. The results of the two
fits were completely consistent, the central value of the 2-mass
fit being slightly lower but the estimated errors larger than 
those of the 1-mass fit. In Table~\ref{rawdata2} we quote only
the outcome of the 1-mass fit. 
Note that the value of $mL$ is in each case large $\gtrsim8$,
and hence finite volume effects are expected to be
smaller than our statistical errors. 
An estimate of the infinite volume second moment mass was then 
computed from the sums $\sum_t S(t)$ and $\sum_t t^2 S(t)$, where for 
$|t|<T_0$ (some large $T_0\le T/2$) we took the measured correlations and 
for $T_0\le |t|<\infty$ we estimated $S(t)$ by assuming it has the 
form of a free lattice correlator 
with the previously determined mass $m$ and amplitude. 
The resulting estimates of $\mr$ from the two fits hardly differed;
in Table~\ref{rawdata2} we quote the results from the 1-mass fit.
The measurements of $\mr$ from $\tilde{G}(k)^{-1}$ were also consistent 
with the values in the table (although statistical errors were 
typically a factor $\sim2$ larger), again indicating small finite 
volume effects.
We also remark that the relations between the quoted $\mr$ and $m$ are 
consistent with estimates from renormalized perturbation theory 
\cite{LWbroken,JMMTW}.

At this stage we reach the goal of determining $\Zr$ through 
Eq.~\eqref{chi}. The corresponding values of $\Zrhat$ are
tabulated in Table~\ref{rawdata2}. For all values of $\kappa$ we find 
$\Zrhat<1$ and consistent with the CW of tending to a non-zero
constant as $\kappa\to\kappac$. There is no signal of a logarithmic
increase as favored by the authors of ref.~\cite{CCC}

What does not agree so well in our measurements with T3 
is the value of $\gr$ for a given $\mr$. As mentioned before,
the estimate of $\gr$ in \cite{LWbroken} is
sensitive to the integration constant $\ln C'_1 =\ln C_1+1/6$;
the estimates in T3 are obtained with 
3--loop RG equations and an input value of $\ln C_1=1.5(2)$.
The present measurements of $\gr$ are better reproduced with
a central value of $\sim1.2$ for $\ln C_1$. 

\subsection{The quantity $v^2\chi$}

In Fig.~1 we plot fits of the data for $v^2\chi$ with functions
\footnote{just adding a constant in the case $p=2$ is of course 
completely ad hoc, but the authors of this paper are actually only 
interested in the case $p=1$}
\be
A_1\ln^p(\kappa-\kappac)+A_2\,,\,\,\,\,\,p=1,2\,.
\end{equation}
To make our point, for the fits we just used the data of ref.~\cite{CCC}.
Our data (the triangles) fall nicely on the fits and illustrate
again the consistency of the data. Both fits are of good quality
with ${\rm chi}^2/{\rm dof}\sim 1$. 
The immediate conclusion is that
the data cannot distinguish between the critical behaviors; to accomplish
this one would have to get much closer to the critical point and treat
much larger correlation lengths. 

Apart from the rather solid theoretical foundation, an additional 
strength of the CW
is that the coefficient of the log ($A_1$) is related to $C'_2$, 
which has been estimated from data in the symmetric phase. 
In fact conventional wisdom predicts a form:
\begin{equation}
v^2\chi=a_1\left[\cl-\frac{25}{27}\ln \cl\right]
+a_2+\rmO\left(\frac{1}{\cl}\right)\,,
\end{equation}
with
\ba
a_1&=&\frac{9C_2^{\prime2}}{32\pi^2}\,,
\\
a_2&=&a_1\left[\ln C'_3+2\ln C'_1+K\right]\,,
\\
K&=&\frac13\ln\left(\frac{3}{16\pi^2}\right)
-\frac{1}{27}\left(7+2\ln2\right)=-1.6317...
\ea
Using the values of the $C'_i$ in \cite{LWbroken} one gets
$a_1=1.20(3)\,,a_2=-1.6(6)$.
Fitting the entire data set with $\kappa\le0.0759$ to the function
(of the RG expected form)
\be
A_1\left[\cl-\frac{25}{27}\ln\cl\right]+A_2\,,
\end{equation} 
we obtain a good fit with ${\rm chi}^2/{\rm dof}=0.9$ and
$A_1=1.267(14)\,,A_2=- 2.89(8)\,,\kappac=0.074833(17)$.  
The fitted value of $\kappac$ is consistent with the best values 
given in Table~\ref{kappac}. Further the $A_i$ are in reasonable 
agreement with the predicted values $a_i$ above.

\begin{table}[ht]
\centering
\begin{tabular}[t]{l|l|l|l|l|l|l}
\hline
$\,\,\,\,\,\kappa$&$L$&$\,\,\,\,\,\,m$&$mL$&$\,\,\,\,\,\mr$
&$\,\,\,\,\,\,\Zrhat$&$\,\,\,\,\,\gr$\\[1.0ex]
\hline \hline
$\kappa_0$&$10$&$0.9642(5)$&$9.6$&$1.0041(5)$ &$0.836(1)$ &$47.94(10)$\\[1.0ex]
$0.077$   &$16$&$0.554(1)$ &$8.9$&$0.563(1)$  &$0.886(3)$ &$36.0(2)$\\[1.0ex]
$0.076$   &$20$&$0.392(1)$ &$7.8$&$0.3950(9)$ &$0.896(4)$ &$30.37(28)$\\[1.0ex]
$0.076^*$ &$20$&$0.390(2)$ &$7.8$&$0.3940(15)$&$0.893(7)$ &$30.08(46)$\\[1.0ex]
$0.0754$  &$32$&$0.266(3)$ &$8.5$&$0.2666(25)$&$0.932(18)$&$27.1(1.0)$\\[1.0ex]
$0.0752$  &$36$&$0.205(2)$ &$7.4$&$0.2054(16)$&$0.901(15)$&$23.06(73)$\\[1.0ex]
$0.0752$  &$48$&$0.205(2)$ &$9.8$&$0.2055(18)$&$0.905(17)$&$23.21(83)$\\[1.0ex]
$0.0751$  &$48$&$0.168(2)$ &$8.1$&$0.1688(15)$&$0.883(17)$&$20.51(74)$\\[1.0ex]
\hline
\end{tabular}
\caption{\footnotesize Measured values of $m,\mr,\Zrhat,\gr$ 
for the Ising model.
In all cases $T=L$ except for the lattice where $\kappa=0.076^*$ denotes
$L=20, T=32$.}
\label{rawdata2}
\end{table}

\begin{figure}
\psfig{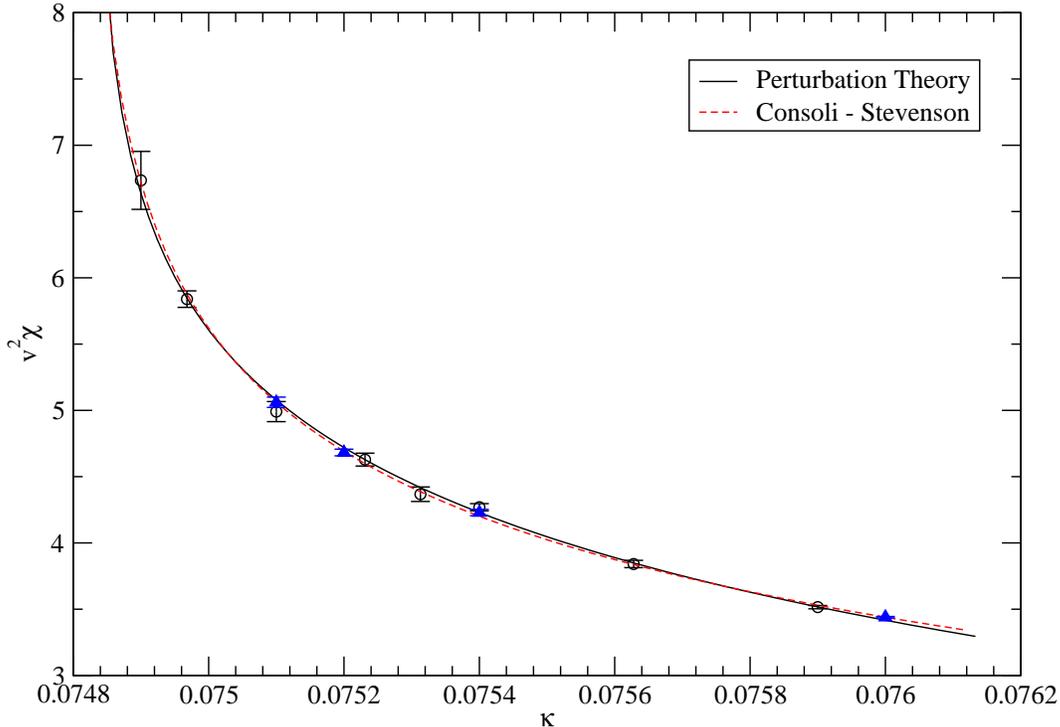}
\caption{\footnotesize Fits of the data (circles) of Cea et al for 
$v^2\chi$. The triangles are the data of this investigation.}
\label{fig1}
\end{figure}

\section{Field theory simulations at constant physics}

We have also studied the cutoff dependence of the renormalization 
constant $\Zr$ along a line of constant IR physics, 
keeping the renormalized coupling
$\gr$ (defined in Eq.~\eqref{gr})
fixed as the renormalized mass $\mr$ in lattice units is varied. 
This is of course the behavior of interest in deciding how limits on
the appearance of new UV physics follows from the triviality of the 
lattice model in the continuum limit. This has been done at three points 
$\bar{\lambda}=0.3, 0.6$ and 1.0 (the Ising limit)
\footnote{for the definition of $\bar{\lambda}$ see \cite{LWsymm}.} 
along the RG curve corresponding to $\gr\simeq20$ 
(cf Fig.~2 of \cite{LWbroken}).
The simulations in the field theory cases ($\bar{\lambda}=0.3,0.6$) 
were done with conventional Metropolis update code on lattices of size 
$64^4$, while the Ising point results were obtained using
a cluster code on lattices of size $48^4$. 

The renormalization constant $\Zr$ and the zero momentum renormalized 
mass $\mr$ were extracted from a fit of the inverse lattice 
propagator in momentum space for small lattice momenta:
\be
\tilde{G}(k)^{-1} = \Zr^{-1}(\mr^{2}+\hat{k}^{2}+\rmO(k^{4}))\,.
\label{eq:propfit}
\end{equation}
In the cases ($\bar{\lambda}=0.3$, $0.6$), we have computed 
$G(k)=\langle|\phi(k)|^{2}\rangle$ from the FFT (fast-Fourier-transform) 
$\phi(k)$ of the coordinate space field $\phi(x)$. The FFT
is performed every 40 Monte-Carlo sweeps: we find that 
the nonzero momentum modes of $\tilde{G}(k)$, for which the vev is 
irrelevant, have an autocorrelation time ranging from 50 to a few 
hundred sweeps. 
For $\bar{\lambda}$=0.3 we have collected propagators for a 
total of 50K sweeps (1250 propagators), while for 
$\bar{\lambda}=$0.6 we have 100K sweeps (2500 propagators). 

Only the lowest $4^4$-1=255 modes are included in the fit:
the mode corresponding to zero momentum is omitted.
We then perform an uncorrelated (diagonal chi-square) fit of the
34 data points (corresponding to different 
values of $\hat{k}^2$) to Eq.~\eqref{eq:propfit}.
The chi-squared/degree of freedom for the $\bar{\lambda}=$0.3 (resp. 0.6)
point was 40/32 (resp. 54/32).
The results for $\Zr,\mr$ obtained by this procedure are indicated
in Table~\ref{phi4res}. Also shown is the field vacuum expectation 
value (unrenormalized) $\langle\phi\rangle$, from which a renormalized 
coupling $\gr$, also shown in the Table, can be obtained.
Finally, we give the susceptibility $\chi$ and the value 
$2\kappa\mr^2\chi$ which is the value of $\Zr$ calculated 
from the zero momentum propagator, Eq.~\eqref{eq:susc}.

It is apparent from the results in Table~\ref{phi4res}
that the cutoff dependence of $\Zr$ is extremely mild over
the entire range covering a factor of 3 in cutoff holding the physical
mass fixed, perfectly in accord with conventional renormalization wisdom.
Note that the values of $2\kappa\mr^2\chi$ 
agree with the latter determination of $\Zr$ within the errors,
as it should be since the Fourier transform of the (subtracted)
correlator, $\tilde{G}(k)$ is a continuous function.
This again shows that the determination of the bare vev is correct.

\begin{table}[htb]
\centering
\vspace{.1in}
\begin{tabular}{c|c|l|l|l} \hline 
 $\bar{\lambda}$ & $\,\,\,\,\,\,\lambda$ & $\,\,\,\,\,\,\,\,\kappa$
  &   $\,\,\,\,\,\,\,\,\,\,v$  &   $\,\,\,\,\,\,\,\mr$ \\[1ex]
\hline
0.3  & 0.275376 & 0.144499 & 0.36895(2)   &  0.5216(48)  \\[1ex]
0.6  & 1.177242 & 0.130307 & 0.13957(13)  &  0.1873(15)  \\[1ex]
1.0  & $\infty$ & 0.0751   & 0.15657(4)   &  0.1691(15)  \\[1ex]      
\hline
\hline
 $\bar{\lambda}$ &  $\,\,\,\,\,\,\,\,\Zrhat$ &$\,\,\,\,\,\gr$ 
&$\,\,\,\,\,\,\,\chi$&$\,\,\,\,\,2\kappa\mr^2\chi$ \\[1ex]
\hline
0.3  &  0.962(12)  & 20.2(4) & \phantom{1}12.40(5) &  0.975(20)  \\[1ex]
0.6  &  0.951(4)   & 19.7(4) & 102.5(3.5)          &  0.937(48)   \\[1ex]  
1.0  &  0.896(8)   & 20.6(7) & 206.4(1.2)          &  0.886(17) \\[1ex]      
\hline
\end{tabular}
\caption{\footnotesize Results for simulations on the RG curve, 
$\gr\simeq20$}
\label{phi4res}
\end{table}

As there may be some question of the extent to which the inverse
momentum space propagator is adequately
fit by Eq.~\eqref{eq:propfit} with two parameters, we have displayed
the quality of the fits for $\bar{\lambda}$=0.3, 0.6
in Figures~2, 3. There is certainly no evidence
for any curvature up to the maximum momenta included in the fit,
and indeed the chi-square/degree of
freedom for the two-parameter fit is perfectly fine.
In Table~\ref{phi4res} for the Ising case we quote numbers
for $\mr$ and $\Zr$, 
obtained by fitting the first three $k \ne 0$ values (with $k$ along the 
time-axis) of the inverse propagator. These numbers differ only slightly
from those given (for the same $\kappa$) in Table~\ref{rawdata2}
using the alternative method of extraction described in the previous 
section. We remark that in this case (as one can check using 
Table~\ref{rawdata0}) the inverse propagator is remarkably linear in 
$\hat{k}^2$ up to the maximal (on-axis) momentum $\hat{k}^2=4$.

\begin{figure}
\psfig{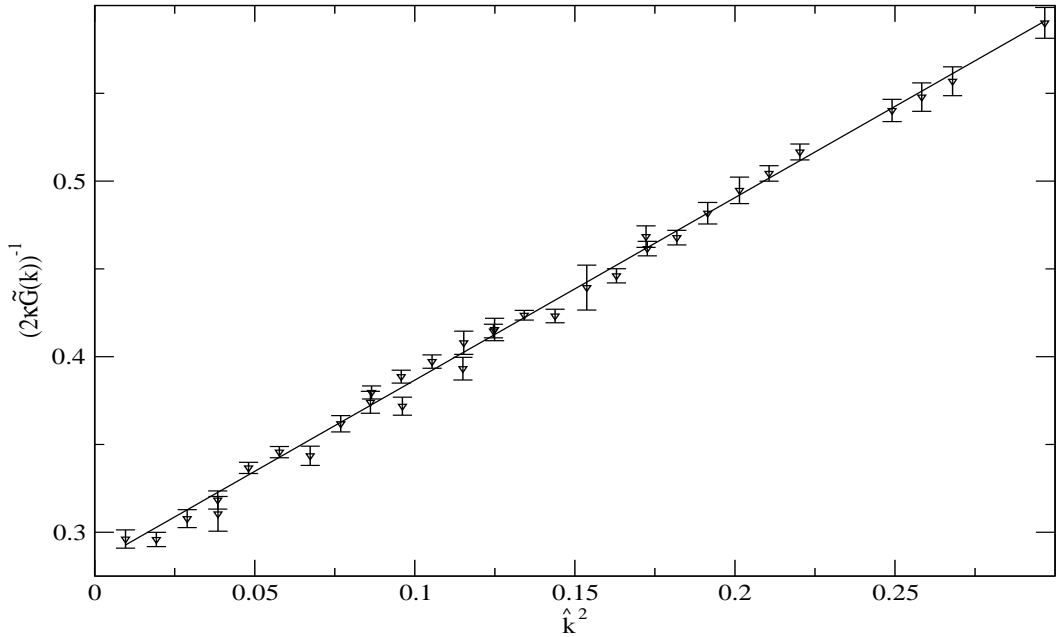}
\caption{\footnotesize $[2\kappa\tilde{G}(k)]^{-1}$ at 
$\bar{\lambda}$=0.3, $\gr\simeq20$. 
The triangles are the data from 50K sweeps 1250 propagators.
The fit is with $\Zrhat,\mr$ given in Table~\ref{phi4res}.}
\label{fig2}
\end{figure}

\begin{figure}
\psfig{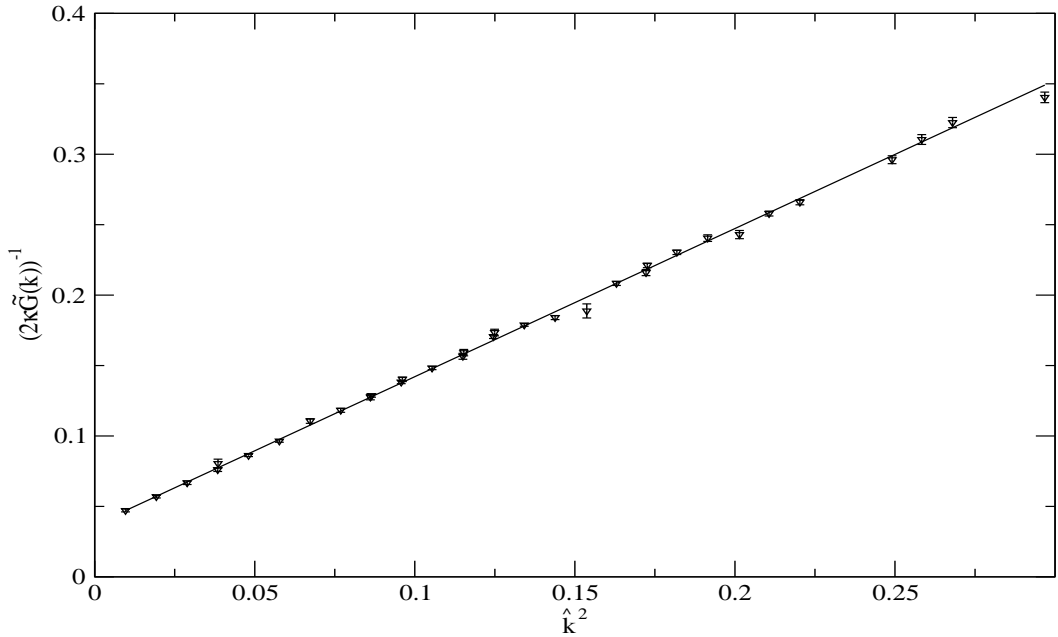}
\caption{\footnotesize $[2\kappa\tilde{G}(k)]^{-1}$ at 
$\bar{\lambda}$=0.6, $\gr\simeq20$. 
The triangles are the data from 100K sweeps 2500 propagators.
The fit is with $\Zrhat,\mr$ given in Table~\ref{phi4res}.}
\label{fig3}
\end{figure}

\section{Conclusions}

In this paper we have shown that simulation data is in perfect 
agreement with the conventional renormalization group scenario
in the lattice 1-component $\phi^4_4$ model. 
This is contrary to the claim of ref.~\cite{CCC}. 
We have explained the deficiencies in their analysis. 
The quantitative agreement with analytically obtained results
is completely satisfactory taking into account that some systematic
errors are probably underestimated in ref.~\cite{LWbroken}.

Conventional wisdom is not firmly established 
merely by a majority vote, but on solid theoretical and
numerical studies. The same applies of course to deeper questions
e.g. whether QCD is the correct theory of hadronic physics.
The work reported here again reinforces the CW regarding triviality.
Had the critique of the authors \cite{CCC} been correct, they would
have indicated serious non-standard implications concerning the Higgs' 
sector of the Standard Model. 

\subsection{Acknowledgments}

We thank the Leibniz-Rechenzentrum where part of the computations
were carried out.
This investigation was supported in part by the Hungarian 
National Science Fund OTKA (under T034299 and T043159)
and by the Schweizerischer Nationalfonds.
The work of A.~Duncan was supported in part by NSF grant PHY-0244599.

\vfill
\eject

% List of references

\eject


\begin{thebibliography}{99}

\bibitem{CCC}
P.~Cea, M.~Consoli, L.~Cosmai,
hep-lat/0407024

\bibitem{LWsymm}
M.~L\"{u}scher, P.~Weisz,
Nucl. Phys. B290 (1987) 25 

\bibitem{LWbroken}
M.~L\"{u}scher, P.~Weisz,
Nucl. Phys. B295 (1988) 65 

\bibitem{BGZ}
E.~Br\'{e}zin, J.~C.~Le Guillou, J.~Zinn-Justin,
``Field theoretical approach to critical phenomena", Vol.6,
Eds. C.~Domb and M.~S.~Green, Academic Press, London (1976)

\bibitem{GSM}
D.~S.~Gaunt, M.~F.~Sykes, S.~McKenzie,
J. Phys. A12 (1979) 871

\bibitem{StaufAd}
D.~Stauffer, J.~Adler, 
Int. J. Mod. Phys. C8 (1997) 263

\bibitem{KL}
R.~Kenna, C.~B.~Lang,
Nucl. Phys. B393 (1993) 461; Erratum-ibid. B411 (1994) 340

\bibitem{VW}
C.~Vohwinkel, P.~Weisz,
Nucl. Phys. B374 (1992) 647

\bibitem{PH} Letter from P.~Hasenfratz to the authors of 
\cite{LWbroken}, Sept.1988

\bibitem{LWO4}
M.~L\"{u}scher, P.~Weisz,
Nucl. Phys. B318 (1989) 705

\bibitem{SW}
R.~H.~Swendsen, J.-S.~Wang, 
Phys. Rev. Lett. 58 (1987) 86

\bibitem{JMMTW}
K.~Jansen, T.~Trappenberg, I.~Montvay, G.~M\"{u}nster, U.~Wolff,
Nucl. Phys. B322 (1989) 698

\bibitem{Binder}
K.~Binder,
Z.  Phys. B43 (1981) 119

\end{thebibliography}
\end{document}